\documentclass[12pt,preprint]{aastex}

\begin{document}

\title{Radio Properties of Low Redshift Broad Line Active Galactic Nuclei}

\author{Stephen E. Rafter, D. Michael Crenshaw, Paul J. Wiita}
\affil{Department of Astronomy and Physics, Georgia State University, Atlanta, Ga. 30303}

\begin{abstract}

\indent
The question as to whether the distribution of radio loudness in active galactic nuclei (AGN) is actually bimodal has been discussed extensively in the literature.  Furthermore, there have been claims that radio loudness depends on black hole mass ($M_{BH}$) and Eddington ratio ($L_{bol}/L_{Edd}$).  We investigate these claims using the low redshift broad line AGN sample of \citet{Greene2007}, which consists of 8434 objects at $z < 0.35$ from the Sloan Digital Sky Survey Fourth Data Release (SDSS DR4).  We obtained radio fluxes from the Very Large Array Faint Images of the Radio Sky at Twenty-Centimeters (FIRST) survey for the SDSS AGN.  Out of the 8434 SDSS AGN, 821 have corresponding observed radio fluxes in the FIRST survey.  We calculated the radio-loudness parameter ($\mathcal{R}$) for all objects above the FIRST detection limit (1 mJy), and an upper limit to $\mathcal{R}$ for the undetected objects.  Using these data, the question of radio bimodality is investigated for different subsets of the total sample.  We find no clear demarcation between the radio loud (RL, $\mathcal{R}$ $>$ 10) and radio quiet (RQ, $\mathcal{R}$ $<$ 10) objects, but instead fill in a more radio-intermediate population in a continuous fashion for all subsamples.  We find that 4.7\% of the AGN in the flux-limited subsample are RL based on core radio emission alone.  We calculate the radio-loud fraction (RLF) as both a function of $M_{BH}$ and $L_{bol}/L_{Edd}$.  The RLF decreases (from 13\% to 2\%) as $L_{bol} / L_{Edd}$ increases over 2.5 orders of magnitude.  The RLF is nearly constant
 ($\sim$5\%) over 4 decades in $M_{BH}$, except for an increase at $M_{BH} > 10^{8} M_{\odot}$.  We find for the FIRST detected subsample that 367 of the RL AGN have $M_{BH} < 10^{8} M_{\odot}$, a large enough number to indicate that RL AGN are not a product of only the most massive black holes in the local universe.

\end{abstract}


\section{Introduction}

\indent
     The nomenclature for classifying active galactic nuclei (AGN) relies on many different factors.  The intrinsic power separates out the lower luminosity Seyfert galaxies from the higher luminosity quasars, whose optical spectra can be very similar.  There is the separation of types 1 and 2 based on the presence or absence (respectively) of broad-lines (BL), an effect presumably caused by the obscuration of the broad-line region (BLR) in type 2 objects due to the presence of apparently toroidally distributed dusty matter in our line of sight.  There is also a variety of different radio properties that yield interesting subcategories.  In general there are the radio quiet (RQ) and the radio loud (RL) populations (based on the ratio of radio to optical flux).  Most RL AGN are found in radio surveys and are further subdivided into Fanaroff-Riley classes (FRI and FRII), where the radio emission from the former mostly arises from the jets and core while the latter are lobe dominated.  There is also a subtle connection to morphology, where most strong emission-line radio galaxies are found in giant elliptical galaxies and most Seyfert galaxies are in spirals \citep{Osterbrock2006,Sikora2007}.  Moving beyond this classification system into one that is more continuous and based on fundamental parameters such as bolometric luminosity ($L_{bol}$, relates directly to accretion rate), black hole mass ($M_{BH}$, relates directly to Eddington luminosity, $L_{Edd}$), Eddington ratio ($L_{bol}/L_{Edd}$, the normalized accretion rate), and inclination (\it i\rm, accounts for the presence or absence of broad-line features) could simplify the subject greatly.  In recent years our ability to determine $M_{BH}$, and hence $L_{bol}/L_{Edd}$, has greatly improved and facilitated this goal.  

     The technique of reverberation-mapping has been used to measure the radii of the broad-line regions (R$_{BLR}$) from which the virial masses of supermassive black holes (SMBHs) can be calculated \citep{Peterson1997book, Peterson2004, Kaspi2000, Kaspi2005} using \begin{equation} M_{BH}=\frac{fR_{BLR}{\Delta}V^2}{G}, \end{equation}  where \it f \rm is a scaling factor that depends on the geometry and kinematics of the BLR and  $\Delta V$ is the velocity dispersion. Empirical relations have been found to relate $R_{BLR}$ to the optical luminosity at 5100 \AA\ ($L_{\lambda}$[5100 \AA]), allowing one simple measurement to be used as a proxy for $R_{BLR}$ \citep{Peterson2004, Kaspi2005}.  There is some debate over the empirical power law that governs this relation, and a few different scalings are used, but recent studies have $R_{BLR} \propto (L_{\lambda}$[5100 \AA])$^\gamma$ where $\gamma \approx 0.5$ \citep{Vestergaard2006}.  This is the (zeroth order) theoretical expectation obtained by assuming a constant ionization parameter (U) and constant electron density ($n_{e}$) for the BLR in all AGN \citep{Peterson1997book}.  The full width at half maximum (FWHM) of the broad-lines gives a measure of the velocity dispersion of the BLR clouds due to their assumed Keplerian orbits around the BH.  These two combined measurements ($L_{\lambda}$[5100 \AA], FWHM) from a single spectrum can give a relatively reliable estimate of $M_{BH}$ \citep{Vestergaard2006}.
     
     It has been long claimed by previous studies that RL AGN make up only about 5\% -- 10\% of the total AGN population \citep{Kellermann1989, Urry1995, Ivezic2002}.  This has led to the speculation that RQ AGN are different from RL AGN because of different physical accretion processes \citep{Ho2002, Sikora2007}.  The idea of two separate populations is based on the claim of a bimodal distribution in the radio-loudness parameter ($\mathcal{R}$) of AGN \citep{Kellermann1999};  $\mathcal{R}$ is simply the ratio of the monochromatic radio luminosity to the monochromatic optical luminosity ($\mathcal{R} \equiv \nu_{radio}L_{radio} / \nu_{opt}L_{opt}$).  Historically, objects are considered to be RL if $\mathcal{R} >$ 10, and RQ if $\mathcal{R} <$ 10 \citep{Kellermann1989}.  Typically the radio luminosity is measured at 5 GHz, and the optical luminosity is measured at 4400 \AA.  One key difference in the current study is that the Very Large Array's (VLA) Faint Images of the Radio Sky at Twenty-centimeters (FIRST) survey operates at 1.4 GHz so the $\mathcal{R}$ value will require a minor scaling.  Corrections to account for different scalings will be discussed in \S2.  Several investigations into this claim of bimodality have yielded results that differ, based on the AGN sample, selection criteria, and inclusion or exclusion of extended radio emission \citep{White2000, Ivezic2002, Ivezic2004, Cirasuolo2003, Cirasuolo2004, Laor2003, White2007}.

     The Sloan Digital Sky Survey (SDSS) allows the optical selection of a large number of AGN with varying types of classifications over a large range in redshifts \citep{Schneider2007}.  Large scale radio surveys like the FIRST survey have allowed characterization of a large number of radio sources at a flux limit of 1 mJy and an angular resolution of 5$''$.  Due to its high resolution, the FIRST survey is primarily sensitive to core emission.  The number of objects with a resolved complex radio morphology (core-jet, core-lobe, double-lobed) in the FIRST survey with an SDSS optical counterpart is less than 10\% \citep{Ivezic2002}, although it is likely that some of the unresolved sources have small, and presumably young, jet or lobe contributions.

     In this paper we take the \citet{Greene2007} sample of low redshift BL AGN from the SDSS, from which $M_{BH}$ and $L_{bol}/L_{Edd}$ can be calculated, and correlate them with their FIRST counterparts to test for bimodality in $\mathcal{R}$ and the dependence of $\mathcal{R}$ on $M_{BH}$.  In this way we can determine the radio properties of AGN in the local universe and compare them with higher redshift samples.  We give a description of the SDSS and FIRST data samples in \S2.  In \S3 we describe our data analysis and biases in the sample.  We present conclusions in \S4.

\section{Data Sample}

\subsection{BL AGN Sample of \citet{Greene2007}}
     The BL AGN sample and all optical data come from \citet{Greene2007} and consists of 8434 AGN, all with z $<$ 0.35 to ensure the observation of H$\alpha$, taken from the SDSS Fourth Data Release (DR4).  Greene \& Ho subtract the stellar continuum from all spectra following their prescription \citep{Greene2004} to ensure that the AGN broad-lines are not masked or suppressed by host galaxy features.  For this study the relevant data taken from the spectra consist of the full width at half maximum of H$\alpha$ (FWHM$_{H\alpha}$) and the luminosity of H$\alpha$ (L$_{H\alpha}$).  With the 35 or so reverberation-mapped AGN, empirical relations between the continuum luminosity and broad H$\beta$ FWHM have been calibrated to derive black hole masses of AGN central engines \citep{Peterson2004}.  Other recent empirical relations have allowed the use of H$\alpha$ to derive M$_{BH}$, because H$\beta$ is usually about 3 times weaker and H$\alpha$ has lower signal-to-noise requirements \citep{Greene2005}.  The following relations were used by \citet{Greene2007} to calculate $M_{BH}$ and $L_{bol}/L_{Edd}$ for all 8434 AGN in their sample:

\begin{equation}
M_{BH} = (2 \times 10^{6}) \left( \frac{L_{{\mathrm H}\alpha}}{10^{42}~{\mathrm {erg~s}}^{-1}}\right)^{0.55} \left( \frac{{\mathrm {FWHM}}_{{\mathrm H}\alpha}}{10^{3}~{\mathrm {km~s}}^{-1}} \right)^{2.06}M_{\odot} 
\end{equation}

\begin{equation}
L_{Edd} = (1.26 \times 10^{38}) \left( \frac{M_{BH}}{M_{\odot}} \right)~{\mathrm {erg~s}}^{-1} .
\end{equation}

To find $L_{bol}/L_{Edd}$, the bolometric luminosity ($L_{bol}$) is required.  \citet{Greene2005} find that $L_{H\alpha}$ scales with $L_{bol}$, which is then calculated using the following relation: 

\begin{equation}
L_{bol} = (2.34 \times 10^{44}) \left( \frac{L_{{\mathrm H}\alpha}}{10^{42}} \right)^{0.86}~{\mathrm {erg~s}}^{-1} .
\end{equation}

The monochromatic continuum luminosity at 5100 \AA\ ($\lambda L_{\lambda}$[5100 \AA]) is found to scale with $L_{bol}$ \citep{McLure2004a, Greene2005}, but because the typical $\mathcal{R}$ value is based on $L_{\lambda}$(4400 \AA), it is necessary to scale $L_{\lambda}$(5100 \AA) using a typical optical spectral index of 0.5 \citep{Sikora2007} so that:

\begin{equation}
\mathrm{
L_{\lambda}(4400 \AA) = (L_{\lambda}[5100 \AA]) \left( \frac{4400 \AA}{5100 \AA} \right)^{-0.5} = (2.59 \times 10^{43}) \left( \frac{L_{H\alpha}}{10^{42}} \right)^{0.86}~erg~s^{-1} }.
\end{equation}

Our final optical data set consists of redshifts, FWHM$_{{\rm H}\alpha}$, $L_{{\rm H}\alpha}$, $L_{\lambda}$(5100 \AA), and $L_{\lambda}$(4400 \AA), along with calculated $M_{BH}$ and $L_{bol}/L_{Edd}$ for all objects in the sample (see Greene \& Ho [2007] for a more complete description of the optical data set).  The full sample has 7565 AGN with $M_{BH} < 10^{8} M_{\odot}$, and 858 AGN with $M_{BH} > 10^{8} M_{\odot}$.  The advantage of using this sample is that it is complete to a given limiting magnitude, and hence a limiting 5100 \AA\ flux ($\sim 1 \times 10^{-17}$ ergs s$^{-1}$ cm$^{-2}$ \AA$^{-1}$), as shown by the solid line in Figure 1. Furthermore this is a spectroscopic sample of broad-line AGN from which $M_{BH}$ and $L_{bol}/L_{Edd}$ can be calculated.  Finally, the AGN in this sample were not, in general, selected on the basis of their radio properties, and thus we can determine the true fraction of AGN at low redshift that are RL.

\subsection{FIRST Data}

     The FIRST survey was designed to cover a large portion of the SDSS coverage area \citep{Becker1995}. The FIRST survey operates at 1.4 GHz with an angular resolution of 5$''$ and has a limiting flux of 1 mJy.  We obtained radio fluxes based on correspondence in position with the optical data.  The search radius was only 4$''$ from the optical counterpart to ensure that the radio emission at the sample redshift limit was still within the host galaxy.  The 4$''$ search radius may, however,  exclude objects with weak core emission and extended radio stuctures such as double lobes or long knotty jets.  Figure 2 shows a histogram of the position offset (indicating the accurate astrometry in both SDSS and FIRST databases).  Only 3\% of the FIRST sources are off by more than 2$''$ from the SDSS optical counterpart (which corresponds to $\sim$14 kpc at the $z = 0.35$ sample limit).  For each object the integrated flux ($F_{int}$) given in mJy was taken.  Out of the 8434 AGN, there were 832 sources with radio flux above the FIRST limit.  Table 1 gives the SDSS and FIRST names of these 832 AGN, along with the side lobe flag indication, integrated flux, and calculated log($\mathcal{R}$) (the table is available in its entirety in the electronic edition).  We removed 11 radio sources with a side lobe flag in the FIRST data due to possible contamination from another nearby source, leaving 821 detected objects.  All objects that did not have a FIRST counterpart were given an upper limit flux of 1 mJy. For consistency, we converted $F_{int}$ into a luminosity using the usual flux-luminosity relation with the same cosmology and redshifts used by \citet{Greene2007} from \citet{Spergel2003} ($H_{0}$ = 71 km s$^{-1}$ Mpc$^{-1}$, $\Omega_{m}$ = 0.27, and $\Omega_{\Lambda}$ = 0.73).  We conform to the standard definition of $\mathcal{R}$ so that comparisons could be made with previous work, where $L_{radio}$ is at 5 GHz.  Since the radio flux in AGN is generally characterized by a power law where $F_{\nu} \propto \nu^{-\alpha}$, the luminosity was scaled by ($\nu_{5GHz}$/$\nu_{1.4GHz}$)$^{-\alpha_{R}} = 0.361$, where the most common radio spectral index, $\alpha_{R}$ = 0.8, is used \citep{Sikora2007}.  So the assumed upper limit assigned to undetected objects of 1 mJy at 1.4 GHz corresponds to 0.361 mJy at 5 GHz.  

     The total sample can be broken into subsets and analyzed separately based on detections and flux limits.  The first subset is the 821 objects that have measured FIRST fluxes.  This subset will be referred to as the `detected' sample.  The detected sample is advantageous in that all $\mathcal{R}$ values will be uniquely determined for each object.  In this subsample, log($\mathcal{R}$) ranges from $-$0.68 to 4.47, spaning over 5 orders of magnitude, and has 683 AGN with $M_{BH} < 10^{8} M_{\odot}$, and 138 AGN with $M_{BH} > 10^{8} M_{\odot}$.  

     The next subset is found by setting an optical flux limit of $F_{\lambda}$(4400 \AA) = 5.6$\times 10^{-17}$ ergs s$^{-1}$ cm$^{-2}$ \AA$^{-1}$ ($F_{\nu}$[4400 \AA] = 0.0361 mJy).  The practical reason for the optical flux limited sample is that for an object not detected by FIRST, we can not uniquely determine if it is RL or RQ if $F_{\nu}$(4400 \AA) $<$ 0.0361 mJy.  All AGN designated RL with an $\mathcal{R}$ upper limit are cut when this optical flux limit is set.  Some of these may be true RL AGN, but since they lack a FIRST counterpart (i.e. too faint for FIRST detection and have a F$_{\nu}$[5 GHz] $<$ 0.361 mJy) they do not make the cut.  Some objects from the detected subsample, and some that are designated RQ with an $\mathcal{R}$ upper limit are cut as well.  This subset will be referred to as the `flux limited' sample and consists of 5477 objects.  The dashed line in figure 1 shows the optical flux limit used to make this cut.  

     The last subset is from the flux limited sample and is found by imposing even more stringent flux limits following \citet{Ivezic2002}.  This last flux imposition takes into account that a large number of objects in a radio vs optical flux plot will tend to overpopulate the region of the plot just above each respective flux limit.  \citet{Ivezic2002} claim that to get a true representation of the $\mathcal{R}$ distribution, this must be taken into account.  Figure 3 is a plot of the radio flux (at 5 GHz) as a function of the optical flux (at 4400 \AA), both given in mJy.  The horizontal and vertical lines indicate the FIRST and SDSS flux limits respectively.  The region just above both of these limits is the most populated, as shown by the dots.  The positively sloped dot-dashed lines are lines of constant log($\mathcal{R}$) $=$ 2,1,0 from the top down, respectively.  According to \citet{Ivezic2002}, it is necessary to take samples in regions that are perpendicular to these lines and well away from the flux limits.  The square symbols are the ones that are in the region of interest.  This `stringently flux limited' sample has 234 objects.

\section{Results}

     In this section we perform our analysis based on different subsamples.  The detected sample is useful when plotting $\mathcal{R}$ as a function of $M_{BH}$ and $L_{bol}/L_{Edd}$, since $\mathcal{R}$ is uniquely determined for all AGN in this subsample.  It is however highly biased when calculating the radio-loud fraction (RLF), in that the number of RQ AGN is significantly underestimated due to the FIRST flux limit.  The flux limited sample is not as affected by this bias since many of the RQ AGN still make the optical flux cut and remain in the subsample, and is therefore used to determine the RLF.  The statistical errors in the RLF plots are calculated from Poisson statistics.

\subsection{Radio-loudness Results}

     Figure 4 shows a histogram of the radio-loudness, $\mathcal{R}$.  The solid line is for all 821 AGN detected with the FIRST survey.  The dashed line in Figure 4 shows the total sample with upper limits on $\mathcal{R}$ calculated for all AGN undetected by FIRST.  The insert shows the peak of the total sample.  In both histograms there is a smooth distribution about log($\mathcal{R}$) $\approx$ 0.9 and no clear evidence for two distinct populations.  The dashed line histogram could always move farther to the left in Figure 4 as indicated by the arrows.  In fact, to actually see a bimodal distribution, the upper limit histogram peak would have to shift very far to the left, with the peak moving at least an order of magnitude or more.  Neither the detected nor upper limit histogram show a double peak profile, indicating a continuous distribution from RL to RQ in this sample.  We note that 51\% of the AGN in the detected sample are RL, using the $\mathcal{R} = 10$ boundary, but this number is extremely biased in that $\sim$90\% of our total sample was below the FIRST detection limit and therefore mostly RQ.

     The $\mathcal{R}$ histogram is shown in figure 5 for the optically flux limited sample.  The solid line is for all AGN above the optical flux limit and detected by the FIRST survey.  The dashed line is for the total sample (undetected FIRST sources included) after the optical flux limit is imposed, with the insert showing the peak.  The sharp cut off at log($\mathcal{R}$) $=$ 1 for the total sample is due to the flux limits imposed.  In both cases there is still no evidence for bimodality in this sample.  The peak for the flux limited histogram is at log($\mathcal{R}$) $\approx$ 0.75.  We find that 4.7\% of the AGN in this subsample are RL.  This percentage is similar to the roughly 5\% - 10\% that have been found in other studies based primarily on higher-redshift AGN \citep{Kellermann1989, Urry1995}.  

     The same plot is shown in Figure 6 for the stringently flux limited sample, with the insert showing the peak of the total sample.  The cutoffs at log($\mathcal{R}$) $=$ 0 and  2 are due to the limits imposed following \citet{Ivezic2002}.  The peak for the stringently flux limited sample is again at log($\mathcal{R}$) $\approx$ 0.75.  Including objects at higher and lower $\mathcal{R}$ values outside the flux cut prescription followed here does not shift the location of the peak or introduce a bimodal distribution with another peak at higher $\mathcal{R}$ values. 

\subsection{Radio-loudness, Eddington Ratio, and Luminosity}

     Figure 7 shows a plot of $\mathcal{R}$ as a function of $L_{bol}/L_{Edd}$ for the detected sample.  There are no AGN with log($L/L_{Edd}$) $< -2.8$ due to the SDSS limiting magnitude.  Although there are broad distributions in both properties, there is a scarcity of AGN in the high accretion rate, high $\mathcal{R}$ regime, and also in the low accretion rate, low $\mathcal{R}$ regime, which is consistent with previous studies that find a global trend of decreasing radio-loudness with increasing Eddington ratio \citep{Ho2002, Nagar2005, Sikora2007}.  Figure 7 does not show two distinct populations of AGN at a given Eddington ratio as found in \citet{Sikora2007}.  The two distinct populations in that paper at intermediate Eddington ratio values (0.01 -- 0.1) likely represent the upper radio-loudness limits of two separate populations, the upper population consisting of BLRGs hosted by giant elliptical galaxies with $M_{BH} > 10^{8} M_{\odot}$, and the lower mostly being the most radio-loud Seyferts, hosted by spiral galaxies, and with $M_{BH} < 10^{8} M_{\odot}$.  Instead we find a continuous distribution from RQ to R,L where the radio intermediate (RI) objects (log($\mathcal{R}$) $\approx$ 1) are the most numerous.

     Figure 8 shows a plot of the 1.4 GHz radio luminosity as a function of $L_{bol}/L_{Edd}$.  The crosses are objects that are RL (defined by $\mathcal{R}$ $>$ 10) and the diamonds are objects that are RQ ($\mathcal{R}$ $<$ 10).  There is significant overlap in the RL/RQ populations around $L_{\nu}$(1.4 GHz) = 10$^{23}$ Watts Hz$^{-1}$, which has previously been used as an alternative division line between RL and RQ sources \citep[e.g.,][]{Best2005}.  We find that 20\% of the objects that have $L_{\nu}$(1.4 GHz) $> 10^{23}$ Watts Hz$^{-1}$ are classified as RQ by $\mathcal{R}$, and that 13\% of these that have $L_{\nu}$(1.4 GHz)$ < 10^{23}$ Watts Hz$^{-1}$ are classified as RL by $\mathcal{R}$.  No obvious trends exist in Figure 8, but it clearly shows the ambiguity that can be involved when these two different definitions of RL and RQ are used.

     In Figure 9, we show the radio-loud fraction (RLF) as a function of log(L$_{bol}$/L$_{Edd}$) for the optically flux limited sample with assumed upper limits.  The RLF is the ratio of the number of RL objects to the total number of AGN, and is plotted in bins of 0.25 dex.  Although the trend is mostly flat over a range of Eddington ratio, the global trend is that the RLF decreases from $\sim$13\% to $\sim$2\% with increasing Eddington ratio, in a manner similar to that found in the previous studies mentioned above.  The stringently flux limited sample shows a similar trend, but with a much higher overall RLF in each bin, due to the flux limits cutting out all of the RQ objects with upper limits on $\mathcal{R}$.

\subsection{Radio-loudness and Black Hole Mass}

     In some studies of AGN at higher luminosities, the degree of radio-loudness was suggested to be dependent on M$_{BH}$, where RL quasars are typically found to have $M_{BH} > 10^{8} M_{\odot}$ \citep{Laor2000, McLure2004b}.  Figure 10 shows a plot of $\mathcal{R}$ vs $M_{BH}$ for our detected subsample.  We find a significant number of RL AGN with $M_{BH} < 10^{8}$ M$_{\odot}$. We find that 367 (53\%) of the objects with $M_{BH} < 10^{8} M_{\odot}$ are RL, compared to 55 (40\%) objects with $M_{BH} > 10^{8} M_{\odot}$ that are RL.  Figure 10 shows that 11 AGN with $M_{BH}  < 10^{8} M_{\odot}$ are extremely RL (log($\mathcal{R}$) $>$ 3).  These outliers do not result from underestimating their H$\alpha$ (and hence continuum) fluxes.  Their radio fluxes are truly high, in the range 136 mJy to 1213 mJy, whereas their H$\alpha$ fluxes have a distribution similar to that of the entire sample.

     Figures 11 \& 12 show the $\mathcal{R}$ distribution for the high and low mass populations for the detected and flux limited subsamples, respectively.  In these figures the solid lines are for the higher mass population and the dashed lines are for the lower mass population.  Both distributions cover the same range and peak in about the same place.  A K-S test shows that two samples drawn from the same population would differ this much 56\% of the time for the detected sample and 84\% of the time for the flux limited sample, indicating that the two distributions are not statistically different.  This is consistent with the findings of \citet{Woo2002} and \citet{Ho2002} that radio-loudness is not intrinsically dependent on $M_{BH}$.  

     Figure 13 shows the RLF as a function of $M_{BH}$, for the optically flux limited subsample.  The RLF here is the ratio of the number of RL AGN to the total number of AGN in mass bins of log($M_{BH}$) = 0.5.  The distribution is essentially flat at the low mass end although there is evidence for an increase in the highest mass bin.

     In Figure 14 we show $L_{\nu}$(1.4 GHz) as a function of $M_{BH}$ with the symbols the same as those in Figure 8.  As in Figure 8 we see a lot of overlap between the RL and RQ AGN, as defined by $\mathcal{R}$ around $L_{\nu}$(1.4 GHz) = 10$^{23}$ Watts Hz$^{-1}$.  This figure is more ordered than Fig.\ 8 in that there are a paucity of points in the low (high) $L_{1.4 {\rm GHz}}$, high (low) $M_{BH}$ range.  Thus, the most powerful radio-core sources tend to have higher black hole masses, but we find little evidence that the radio-loudness depends strongly on mass.

\section{Conclusions}
     
     We have taken 8434 optically selected BL AGN from the SDSS and estimated $M_{BH}$ and $L_{bol}/L_{Edd}$ for them.  Using fluxes taken from the FIRST survey we compute $\mathcal{R}$, the radio-loudness parameter, where the radio emission is thought to be mostly from the core.  We note that because of the resolution of the FIRST survey and our small search radius (4$''$), we miss AGN with extended radio structures that have very weak core emission.  We note that $\sim$90\% of all SDSS/FIRST matched AGN have only core radio emission \citep{Ivezic2002}, although because of the resolution limit of FIRST, many of these `core' sources will include compact jets and lobes.  Furthermore, the higher redshift samples with which we are comparing our results \citep{White2000, Ivezic2002, Cirasuolo2004} are also based on core radio emission from FIRST.  In the flux-limited subsample we find that 4.7\% of the AGN are RL.  This is similar to the approximately 5\% found for higher redshift quasars, but may be coincidental in that the distribution of radio-loudness likely depends on the available range of Eddington ratio and host galaxy morphology in the sample.  The issue of extended radio structure and its impact on the radio-loudness distribution will be addressed in a forthcoming paper.   
     
     The radio-loudness histograms in Figures 4 \& 5 both peak in the same place, at log($\mathcal{R}$) $\approx$ 0.75.  We find no evidence for a bimodal histogram for any subset of the total sample.  We do note that the upper limit (dashed-line) histograms could be pushed farther to the left to reveal a bimodal distribution, but at much lower $\mathcal{R}$ values than previously found by \citet{Ivezic2004} and \citet{Cirasuolo2004}, who find peaks at log($\mathcal{R}$) $\ll$ 1, and log($\mathcal{R}$) $\approx$ 2-3.  After we apply various flux limits to our total sample, we still cannot reproduce a local minimum at log($\mathcal{R}$) $\approx$ 1.  In all our subsamples we fill in a substantial number of radio-intermediate (RI) objects.  The minimum in the $\mathcal{R}$ histogram could be shifted to lower values or could just not be present at all at these low redshifts.  Nevertheless, we still see a RL tail.  Compared to higher redshift samples, the $\mathcal{R}$ distribution in the local universe for detected sources is narrower and shifted to lower values \citep{Ivezic2004, White2007}.  This may indicate that the radio luminosity function has a stronger evolution than the optical luminosity function to the current epoch, as suggested by \citet{Jiang2007}.  \citet{White2007} find for their sample a similar effect of evolution of the $\mathcal{R}$ histogram with redshift.  Our plot is in good agreement with their low-redshift ($z < 0.5$) quasars, where there is a peak at low values of $\mathcal{R}$, and a tail that falls off over 4 orders of magnitude, with no indication of two separate populations.    

     Our plot of $\mathcal{R}$ as a function of $L_{bol} / L_{Edd}$ lacks points in the low accretion/RQ and high accretion/RL regimes.  We also see a trend of higher RLF at the low values of $L_{bol} / L_{Edd}$.  This matches the general trend seen by \citet{Ho2002} and \citet{Sikora2007} of decreasing $\mathcal{R}$ with increasing L$_{bol}$/L$_{Edd}$, based on AGN that extend to much lower L$_{bol}$/L$_{Edd}$ ($\sim$10$^{-7}$).
     
     We do not find two separate populations at a given $L_{bol}/L_{Edd}$, but again fill in a more RI population.  This may be due to the homogeneous optical selection of our data set, as opposed to the admittedly heterogeneous selection criteria of AGN in \citet{Sikora2007}, which includes AGN from various samples, where objects were selected in very different ways.  We do note however that we do not cover nearly as much parameter space as Sikora et al.\ who get up to an $\mathcal{R}$ of $\sim 10^{7}$ and down to an $L_{bol}/L_{Edd}$ of $\sim$ 10$^{-7}$.  Our sample does not go down to the very low $L_{bol}/L_{Edd}$ values due to the SDSS flux limit.  Homogeneous samples like ours that extend to much fainter radio and optical fluxes would be extremely useful for further studies.  In fact, going deeper in both the radio and optical is necessary if one wants to search for AGN in the very lowest $L_{bol}/L_{Edd}$ and $\mathcal{R}$ regimes. 

     We find a substantial number of RL AGN with $M_{BH} < 10^{8} M_{\odot}$, as seen in Figure 10 (although the fraction of RL sources is largest for the most massive BHs in our sample as shown in Fig.\ 13).  This clearly shows that in the local universe, not only massive BHs ($M_{BH} > 10^{8} M_{\odot}$) are responsible for RL AGN, which is consistent with the findings of \citet{Ho2002} and \citet{Woo2002}.

\acknowledgments
We thank Jenny Greene and Luis Ho for providing their data set and for helpful advice.  We appreciate conversations with {\v Z}eljko Ivezi{\'c} and Amy Kimball.  We thank the anonymous referee for helpful suggestions.  
PJW was supported in part by a subcontract to GSU from NSF grant AST05-07529
to the University of Washington.  Funding for the SDSS Archive has been provided by the Alfred P. Sloan Foundation, the Participating Institutions, the National Aeronautics and Space Administration, the National Science Foundation, the U.S. Department of Energy, the Japanese Monbukagakusho, and the Max Planck Society. The SDSS Web site is http://www.sdss.org/.  The SDSS is managed by the Astrophysical Research Consortium (ARC) for the Participating Institutions. The Participating Institutions are The University of Chicago, Fermilab, the Institute for Advanced Study, the Japan Participation Group, The Johns Hopkins University, Los Alamos National Laboratory,the Max-Planck-Institute for Astronomy (MPIA), the Max-Planck-Institute for Astrophysics (MPA), New Mexico State University, Princeton University, the United States Naval Observatory, and the University of Washington.  The FIRST Survey is supported by grants from the National Science Foundation, NATO, the National Geographic Society, Sun Microsystems, and Columbia University.

\begin{deluxetable}{cccrr}
\tablecolumns{5}
\tabletypesize{\small}
\tablewidth{0pc}
\tablecaption{The Matched SDSS and FIRST Sample\label{table}}
\tablehead{
\colhead{SDSS Name} & \colhead{FIRST Name} & \colhead{Flag} &
\colhead{$F_{int}$} & \colhead{log($\mathcal{R}$)} \\
\colhead{(1)} & \colhead{(2)} & \colhead{(3)} & \colhead{(4)} &
\colhead{(5)}
}

\startdata
J000202.95$-$103037.9 & 00015$-$10405G &   0 &      2.21 &  0.86 \\
J000710.01$+$005329.0 & 00075$+$01050E &   0 &      1.44 &  0.27 \\
J000813.22$-$005753.3 & 00075$-$01050E &   1 &      2.74 &  0.94 \\
J002141.01$+$003841.8 & 00225$+$00390E &   0 &      1.38 &  1.17 \\

\enddata
\tablecomments{
Col.(1): SDSS Name;
Col.(2): FIRST Name;
Col.(3): Flag (1 $=$ bad flag);
Col.(4): Integrated Radio Flux at 1.4 GHz (mJy);
Col.(5): Radio Loudness Parameter.
}
\end{deluxetable}
     
\begin{figure}[b]
\includegraphics[angle=90, scale=0.6]{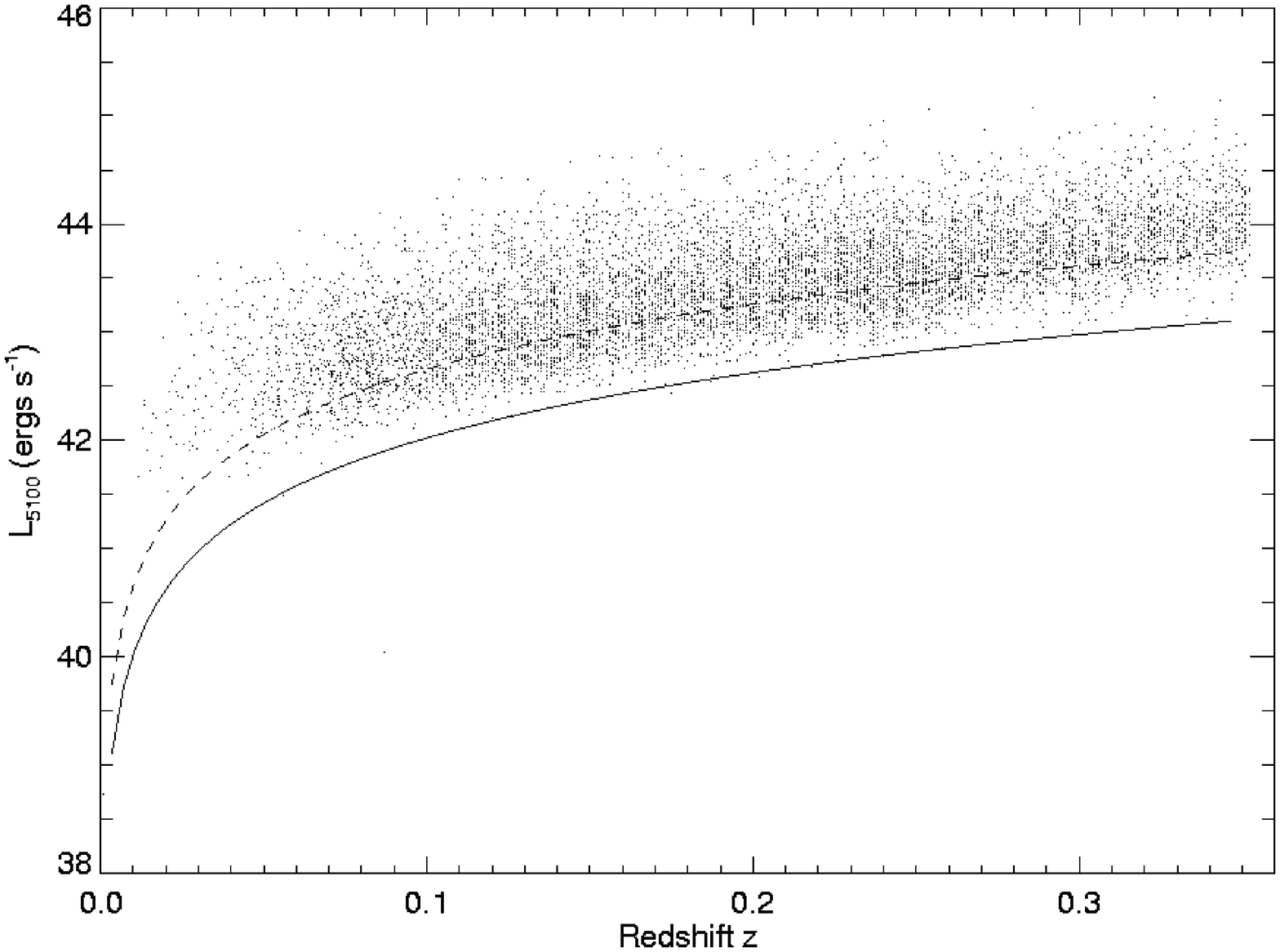}
\caption[]{}
$\lambda$L$_{\lambda}$(5100 \AA) vs z for all SDSS AGN.  The solid line is an approximate SDSS flux limit (F$_{\lambda}$(5100 \AA) $=$ 1$\times$10$^{-17}$ ergs s$^{-1}$ cm$^{-2}$ \AA$^{-1}$).  The dashed line indicates the flux limit of the flux limited sample (F$_{\lambda}$(5100 \AA) $=$ 3.9$\times$10$^{-17}$ ergs s$^{-1}$ cm$^{-2}$ \AA$^{-1}$).
\label{fig1}
\end{figure}

\begin{figure}[t]
\includegraphics[angle=90, scale=0.6]{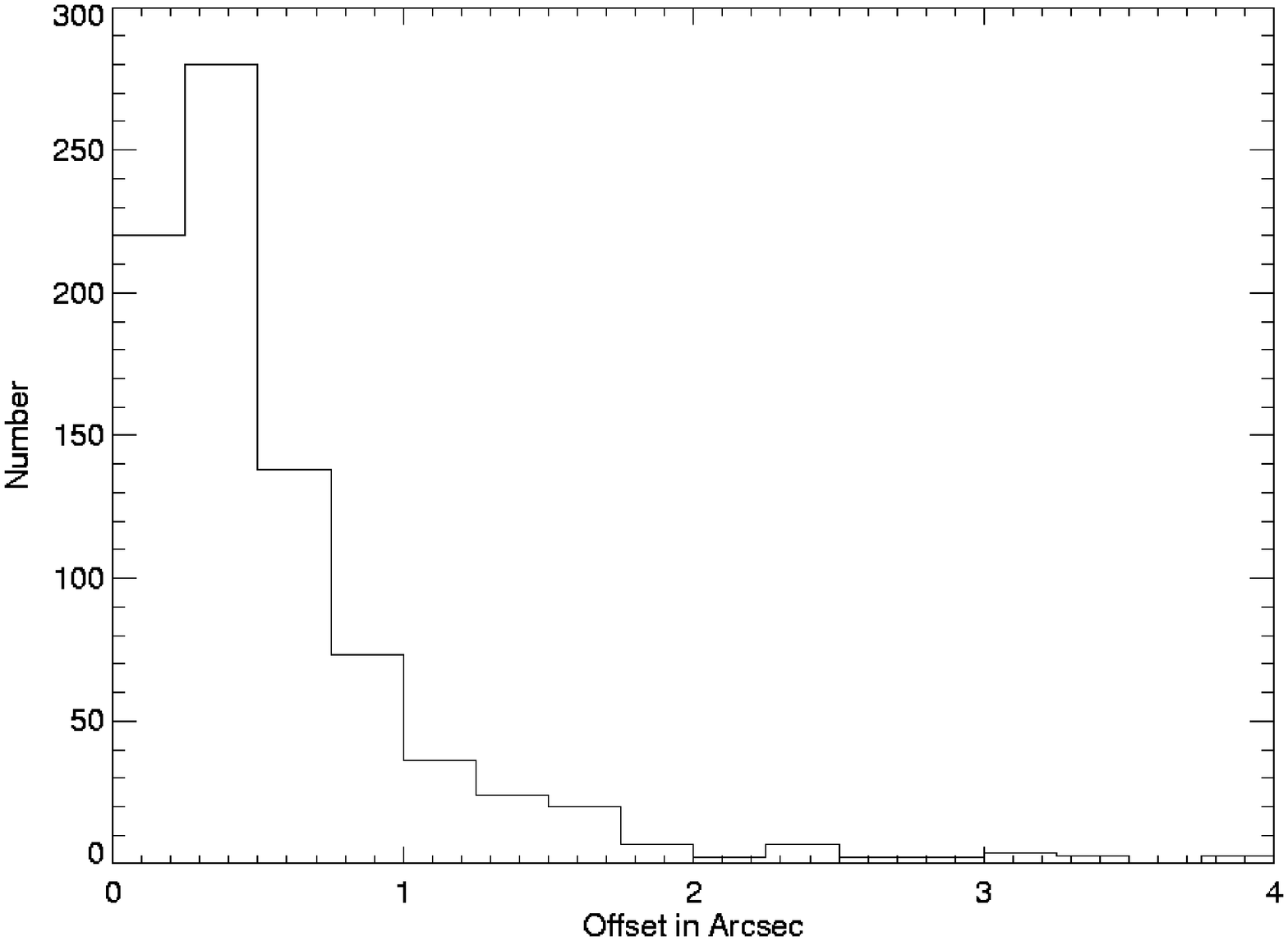}
\caption[]{}
Position offset in arcsecs.  Only 3\% are off by more than 2$''$.
\label{fig2}
\end{figure}

\begin{figure}[t]
\includegraphics[angle=0, scale=0.8]{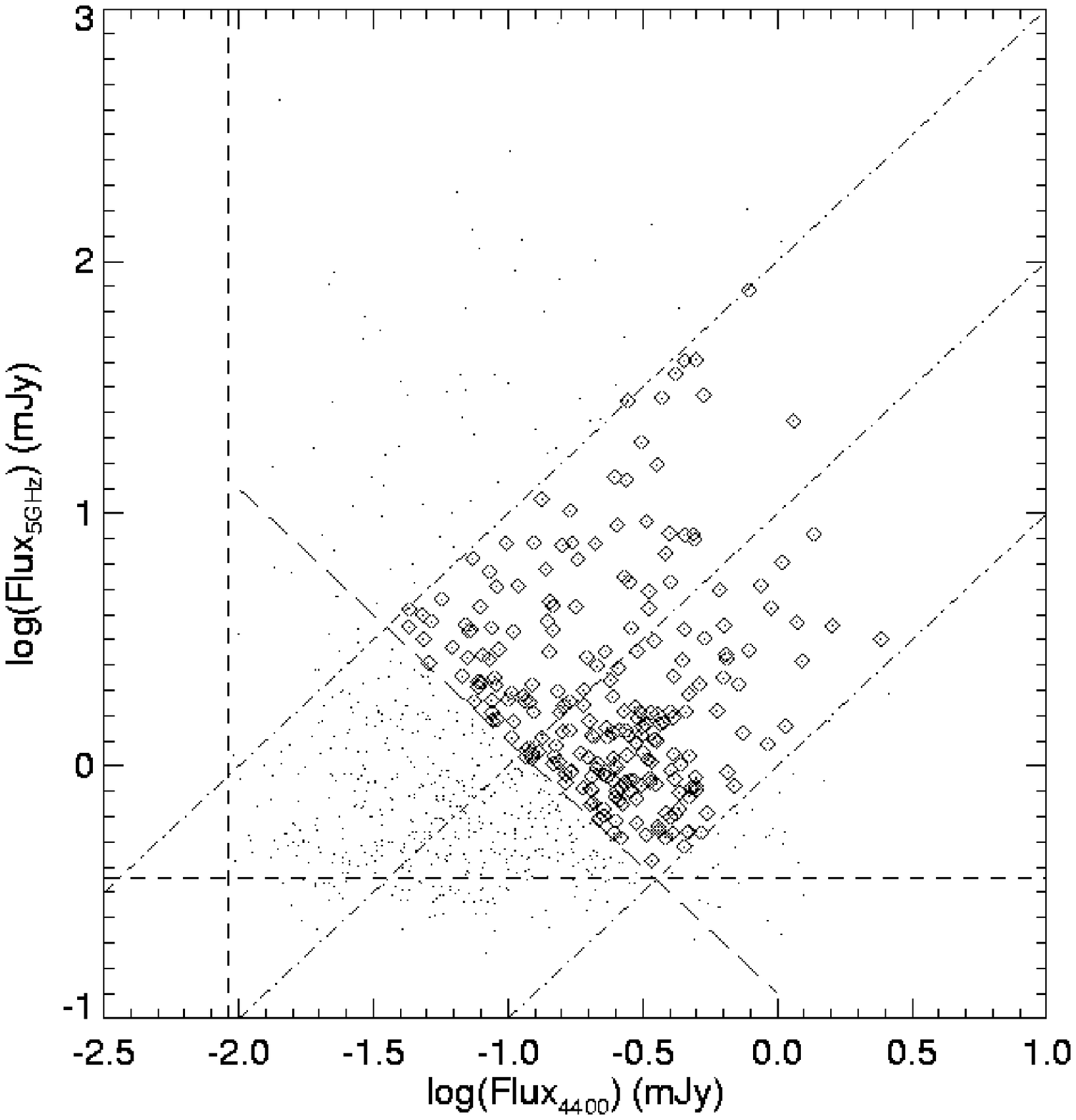}
\caption[]{}
5 GHz radio flux vs. 4400 \AA\ optical flux.  The horizontal and vertical dashed lines indicate the FIRST and (approximate) SDSS flux limits respectively.  The positively sloped dot-dashed lines are the lines of constant log($\mathcal{R}$) = 2, 1, 0 (from the top line down).  Objects above the line perpendicular to the log($\mathcal{R}$) lines defines a boundary for the stringently flux limited sample.  The dots are objects that are outside the region of interest and do not make this sample; the squares are objects that do satisfy these flux cuts.
\label{fig3}
\end{figure}

\begin{figure}[t]
\includegraphics[angle=90, scale=0.6]{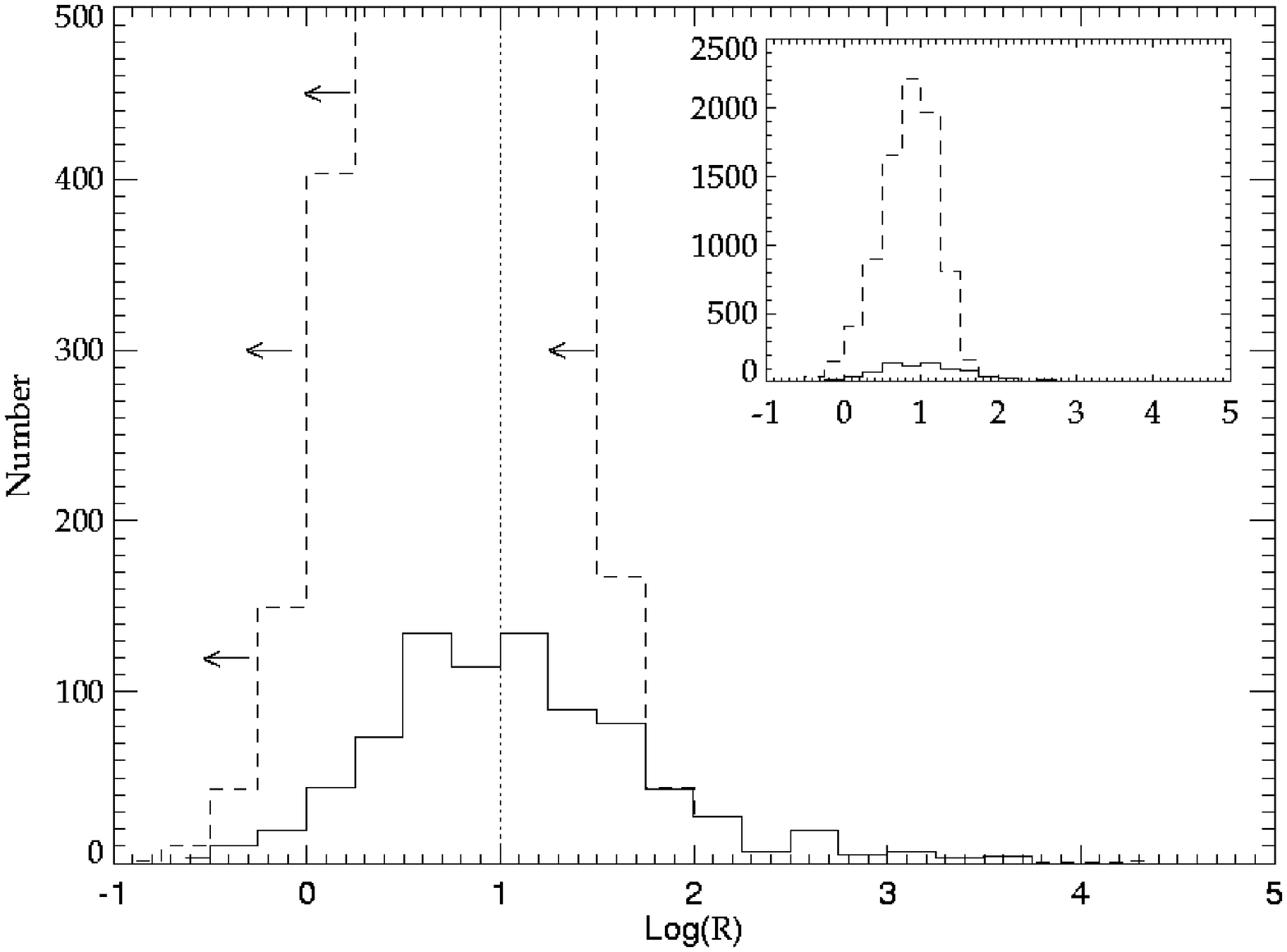}
\caption[]{}
Radio-loudness ($\mathcal{R}$) histogram:  the solid line is for the 821 objects in the detected subsample and the dashed line is a cumulative upper limit for the entire sample using a FIRST flux limit of 1 mJy at 5 GHz.  The insert shows the peak of the upper limit histogram.  Both peak at log($\mathcal{R}$) $\approx$ 0.9.
\label{fig4}
\end{figure}

\begin{figure}[t]
\includegraphics[angle=90, scale=0.6]{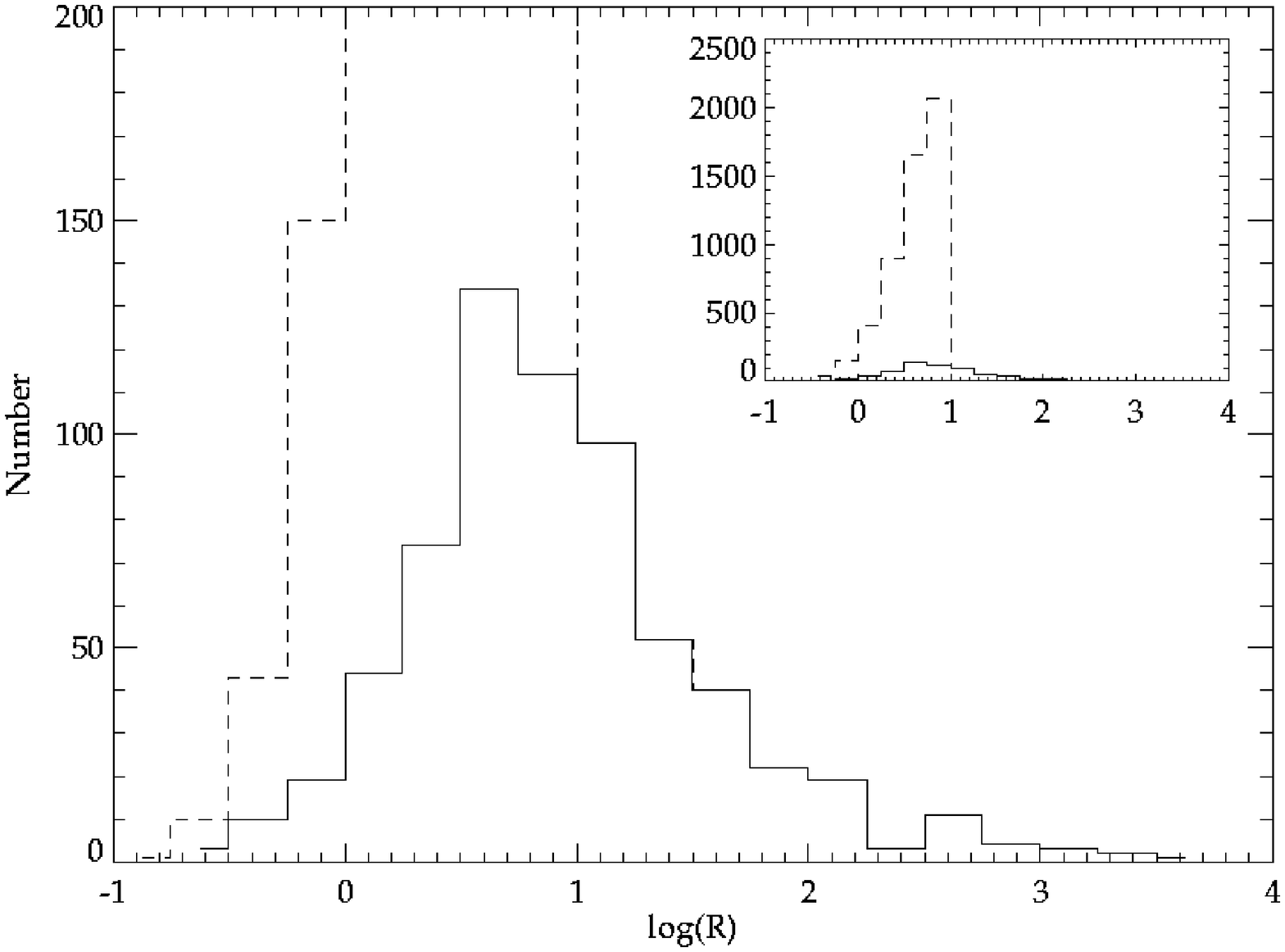}
\caption[]{}
Flux limited Radio Loudness ($\mathcal{R}$) histogram:  the solid line is for the AGN in the detected subsample after the optical flux limit is imposed and the dashed line is an upper limit for all AGN with the optical flux limit imposed.  The insert shows the peak of the upper limit histogram. 
\label{fig5}
\end{figure}

\begin{figure}[t]
\includegraphics[angle=90, scale=0.6]{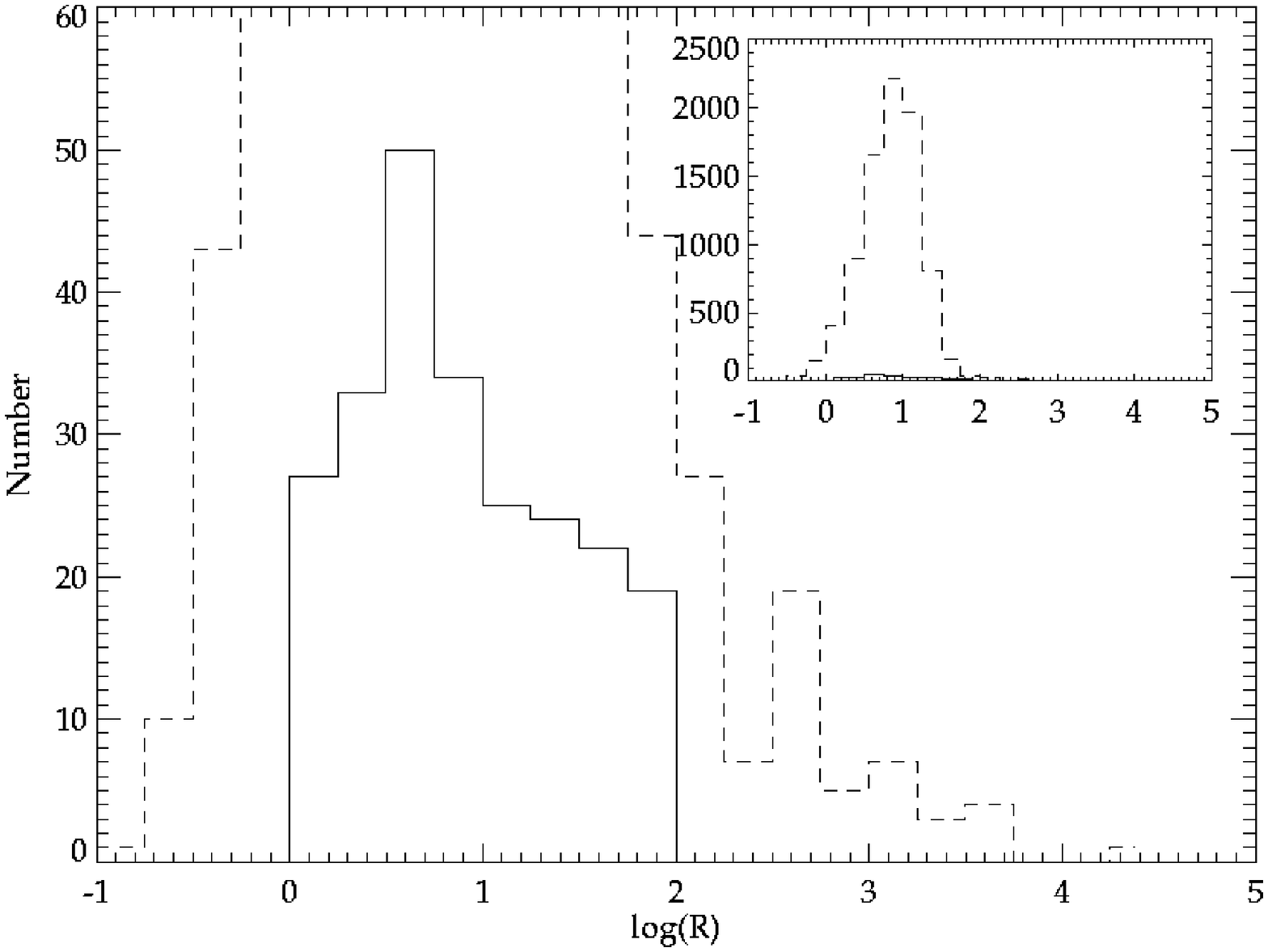}
\caption[]{}
Stringently flux limited $\mathcal{R}$ histogram:  the solid line is for the AGN in the stringently flux limited subsample and the dashed line is the upper limit from the total sample (same as in Figure 4).  The insert shows the peak of the upper limit histogram.
\label{fig6}
\end{figure}

\begin{figure}[t]
\includegraphics[angle=90, scale=0.6]{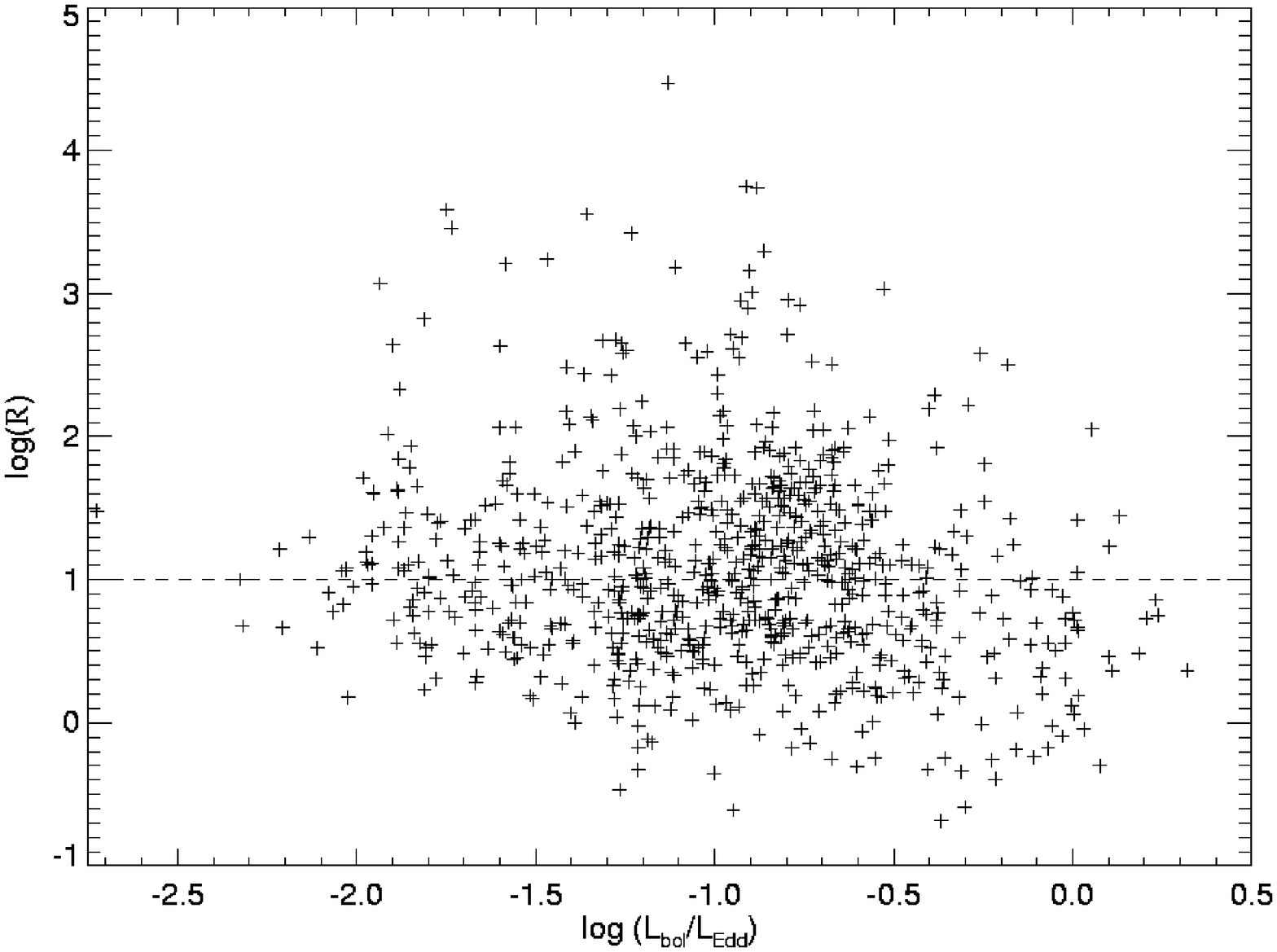}
\caption[]{}
$\mathcal{R}$ vs $L_{bol} / L_{Edd}$.  The dashed line shows the division between RL and RQ.  There are clearly a lack of points in both the lower left and upper right of this plot, indicating the paucity of RL AGN that have high accretion rates and of RQ AGN that have low accretion rates.
\label{fig7}
\end{figure}

\begin{figure}[t]
\includegraphics[angle=90, scale=0.6]{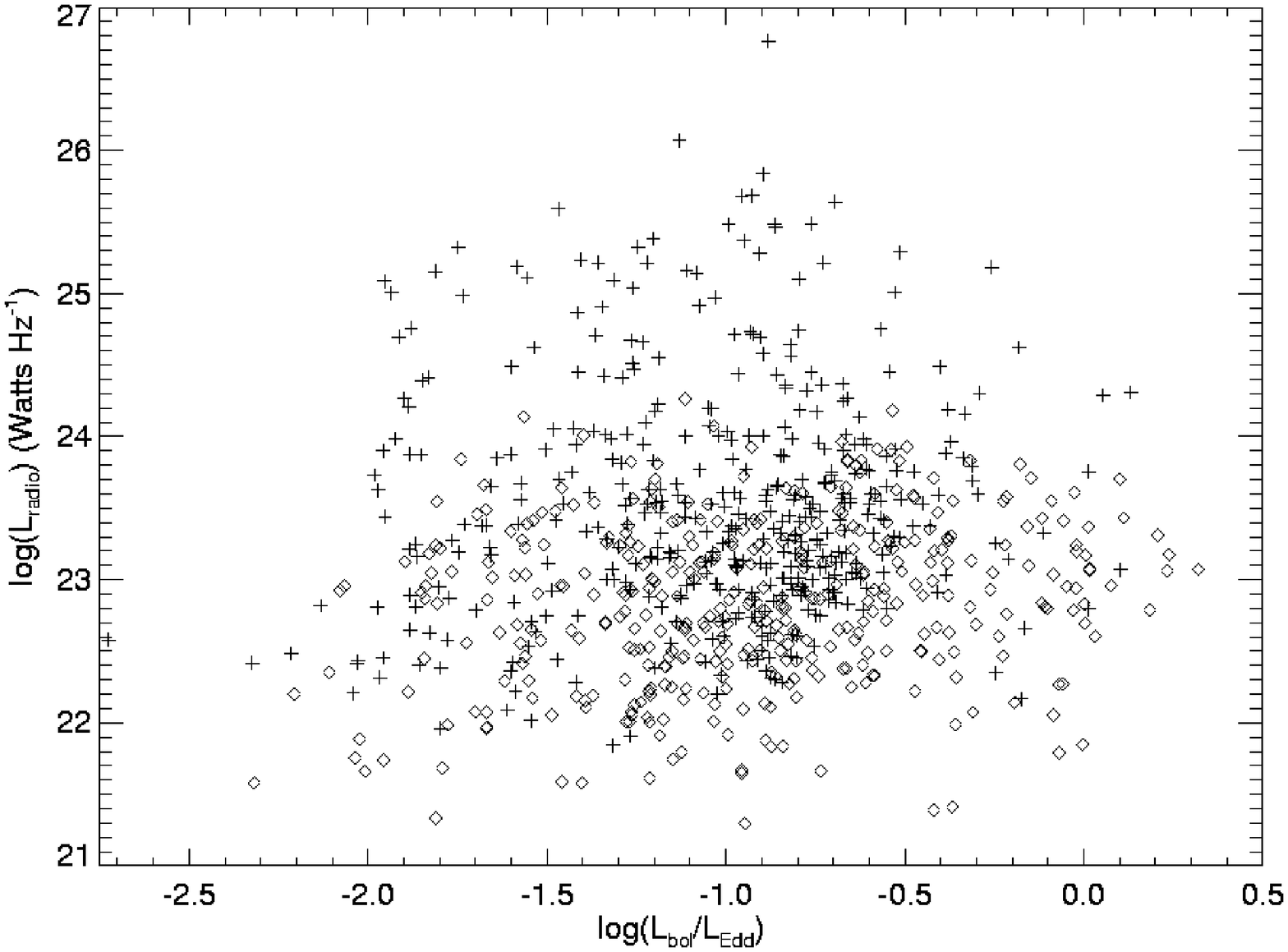}
\caption[]{}
$L_{1.4 {\rm GHz}}$ vs $L_{bol}/L_{Edd}$.  The crosses have $\mathcal{R} > 10$ and the diamonds have $\mathcal{R} < 10$.
\label{fig8}
\end{figure}

\begin{figure}[t]
\includegraphics[angle=90, scale=0.6]{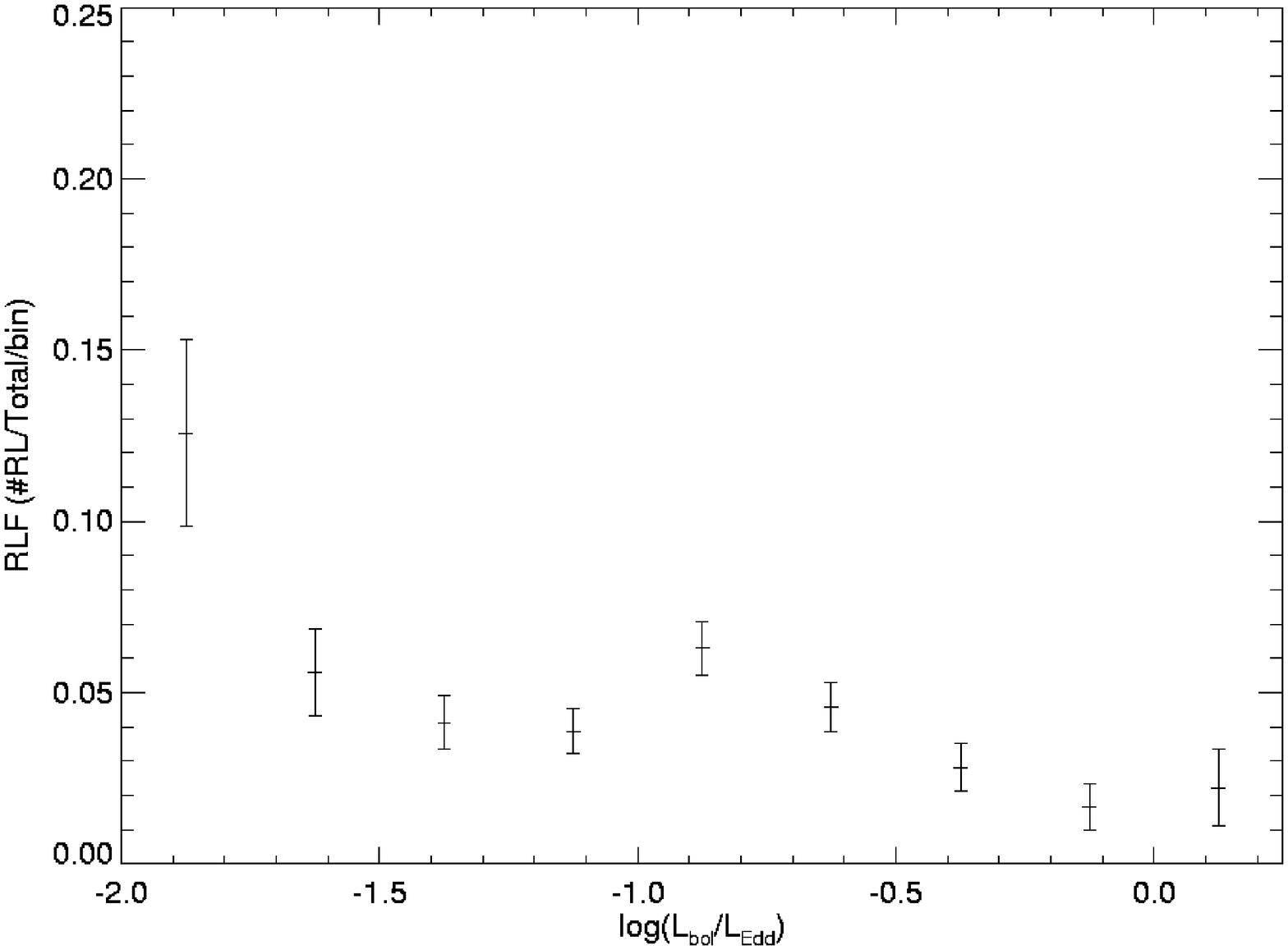}
\caption[]{}
Radio-loud fraction (RLF) for the optically flux limited sample in bins of 0.25 dex.  Error bars are found from a Poisson distribution reduced $\chi$$^{2}$ procedure.
\label{fig9}
\end{figure}

\begin{figure}[t]
\includegraphics[angle=90, scale=0.6]{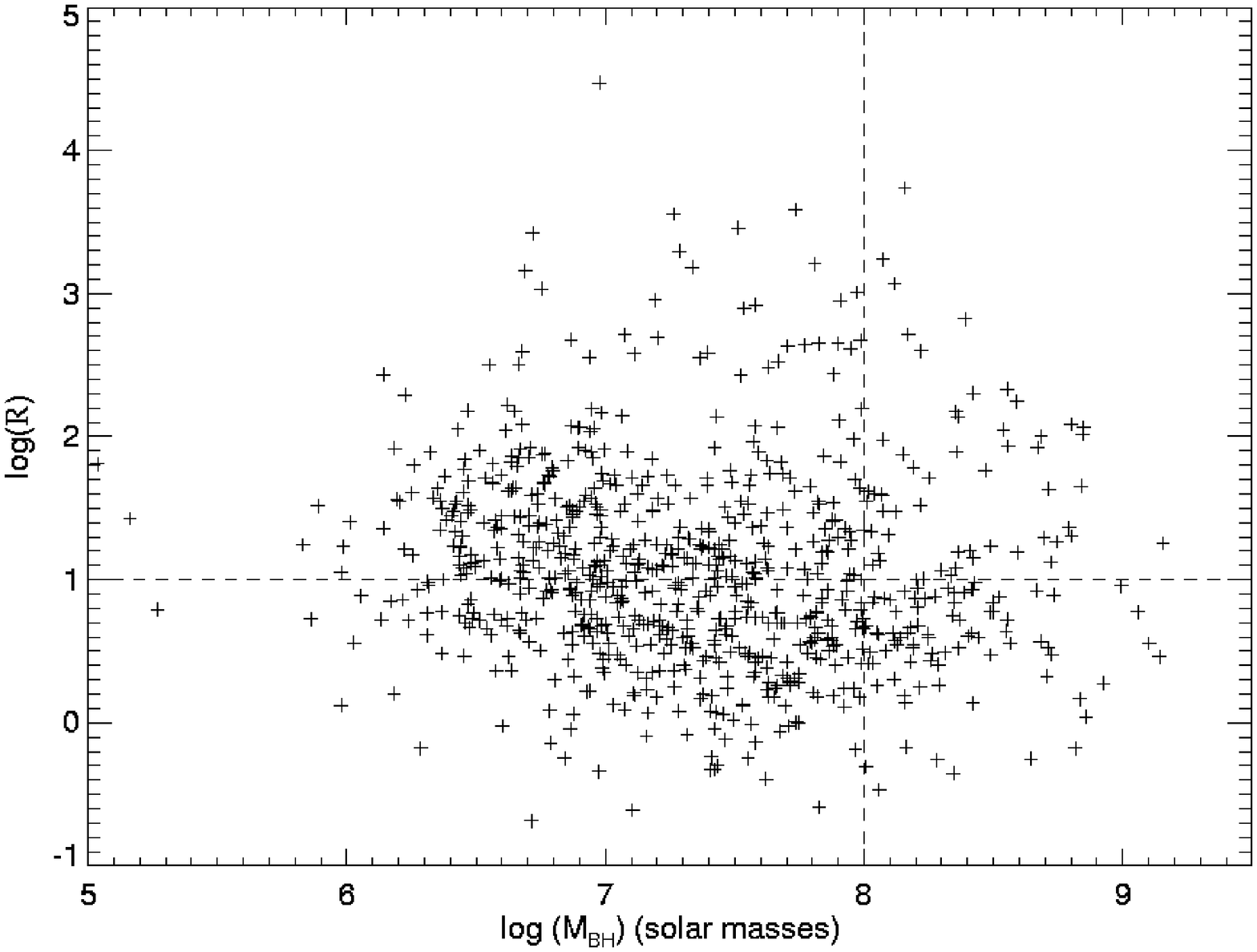}
\caption[]{}
$\mathcal{R}$ vs M$_{BH}$:  The horizontal dashed line is the division between RL and RQ.  The vertical line is the `conventional' separation line between more massive ($M_{BH} > 10^{8} M_{\odot}$) and less massive ($M_{BH} < 10^{8} M_{\odot}$) black holes.  We find 367 AGN with $M_{BH} < 10^{8} M_{\odot}$.
\label{fig10}
\end{figure}

\begin{figure}[t]
\includegraphics[angle=90, scale=0.6]{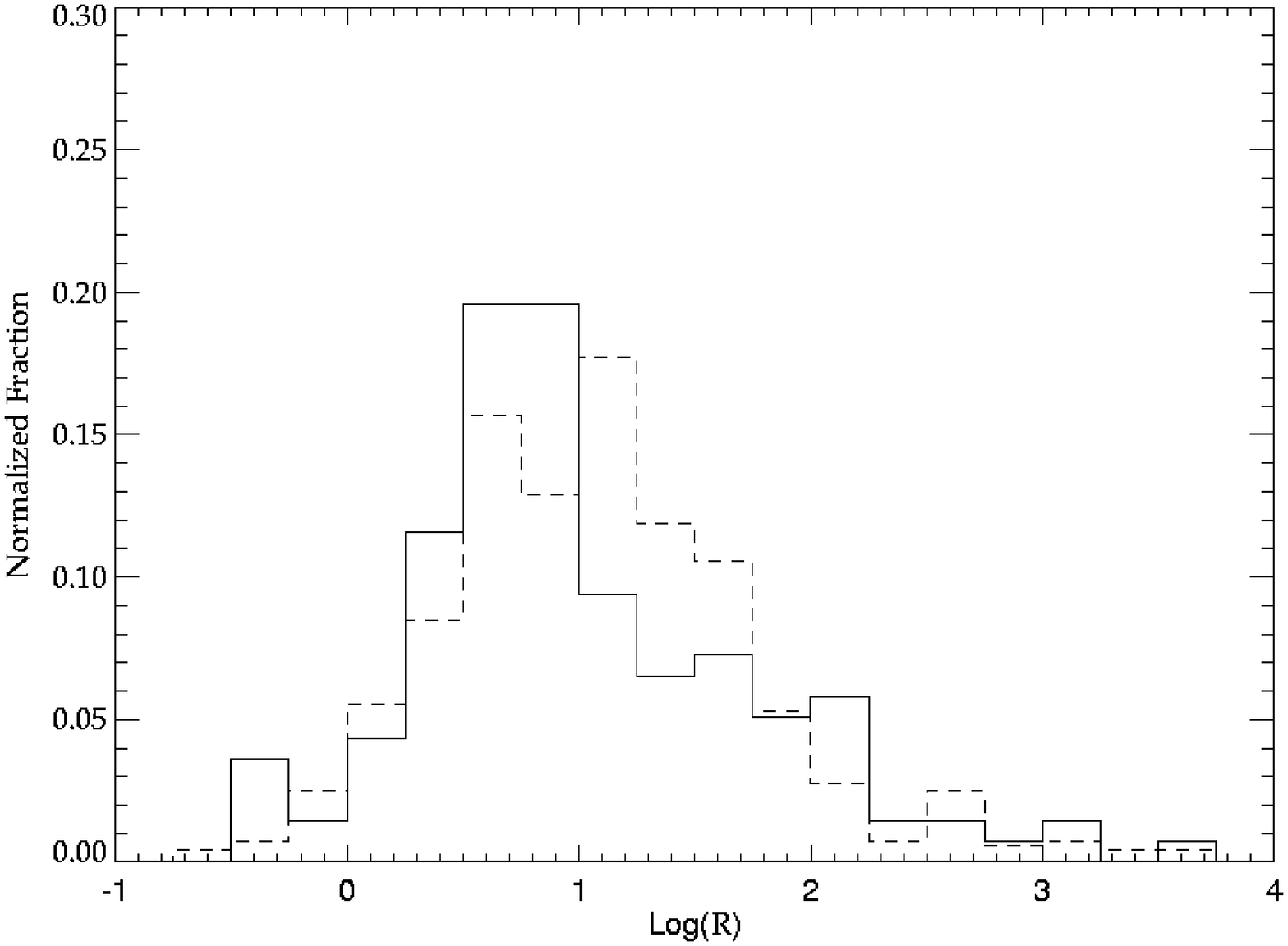}
\caption[]{}
Histogram of the $\mathcal{R}$ distribution of the high mass ($M_{BH} > 10^{8} M_{\odot}$, solid line) and the low mass ($M_{BH} < 10^{8} M_{\odot}$, dashed line) AGN for the detected sample. The two populations have a K-S value of 0.555.
\label{fig11}
\end{figure}

\begin{figure}[t]
\includegraphics[angle=90, scale=0.6]{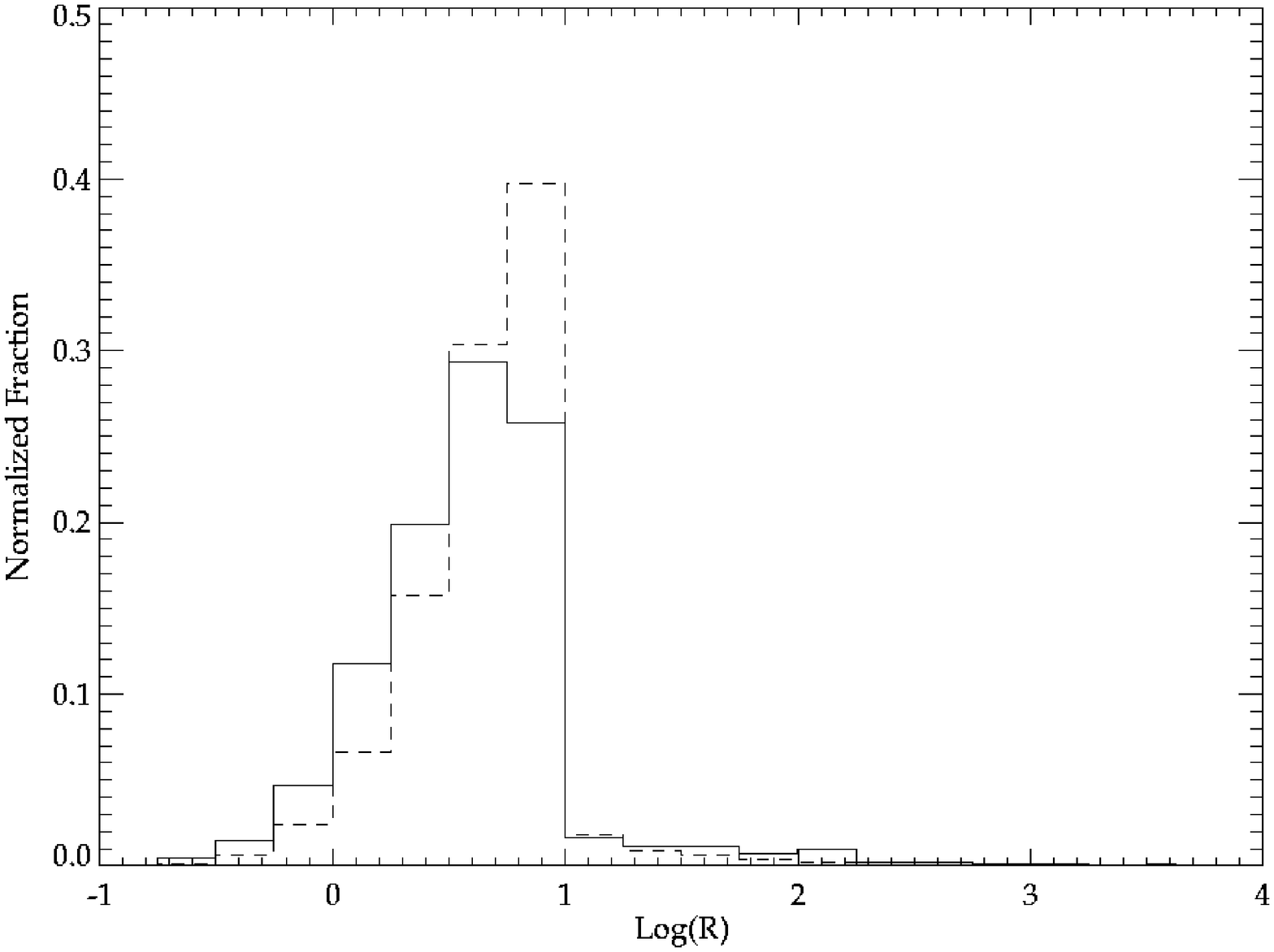}
\caption[]{}
Histogram of the $\mathcal{R}$ distribution of the high mass ($M_{BH} > 10^{8} M_{\odot}$, solid line) and the low mass ($M_{BH} < 10^{8} M_{\odot}$, dashed line) AGN for the flux limited sample.  The two populations have a K-S value of 0.841.
\label{fig12}
\end{figure}

\begin{figure}[t]
\includegraphics[angle=90, scale=0.6]{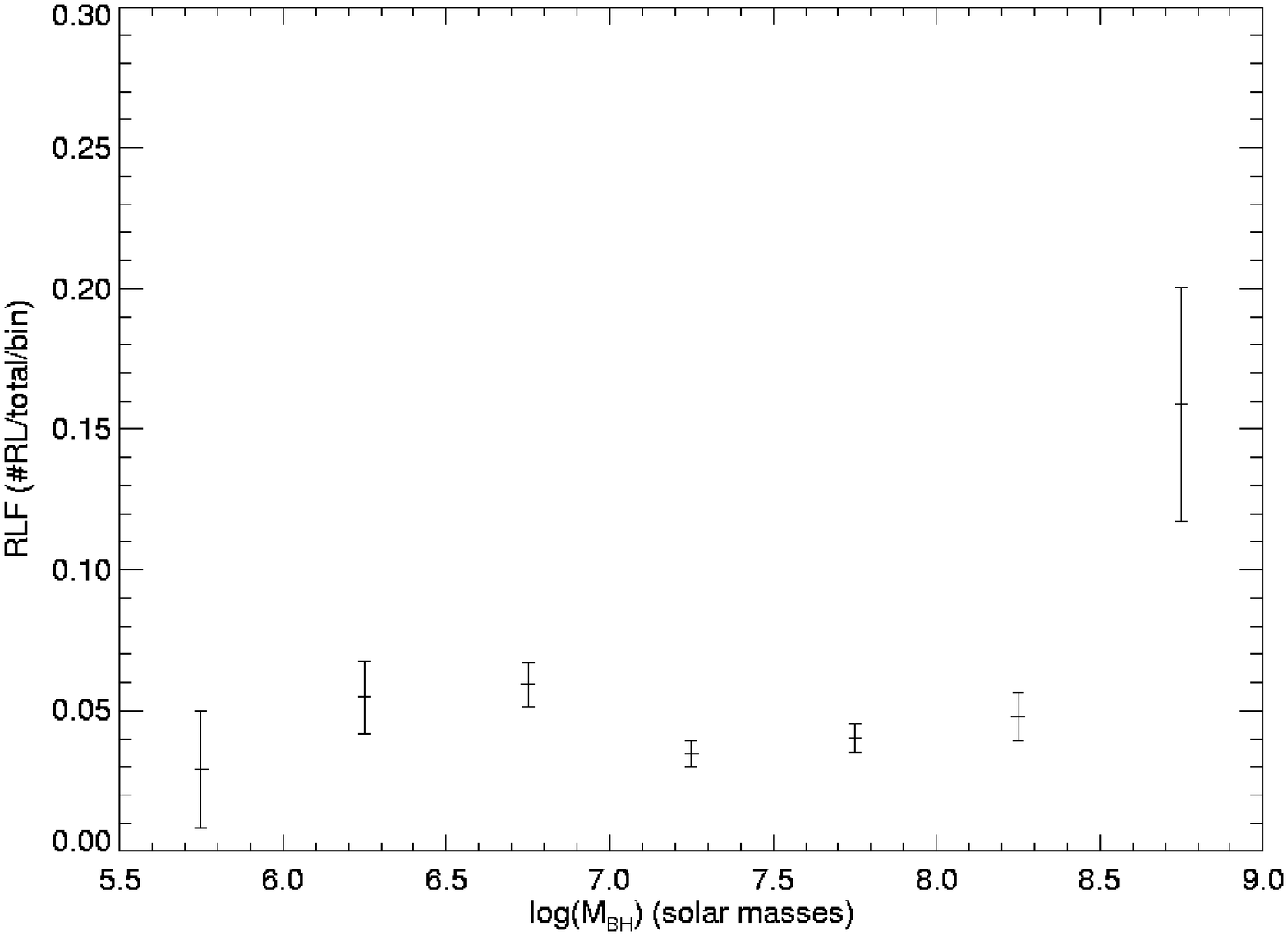}
\caption[]{}
Radio-loud fraction (RLF) for the optically flux limited sample in bins of 0.5 dex.
\label{fig13}
\end{figure}

\begin{figure}[t]
\includegraphics[angle=90, scale=0.6]{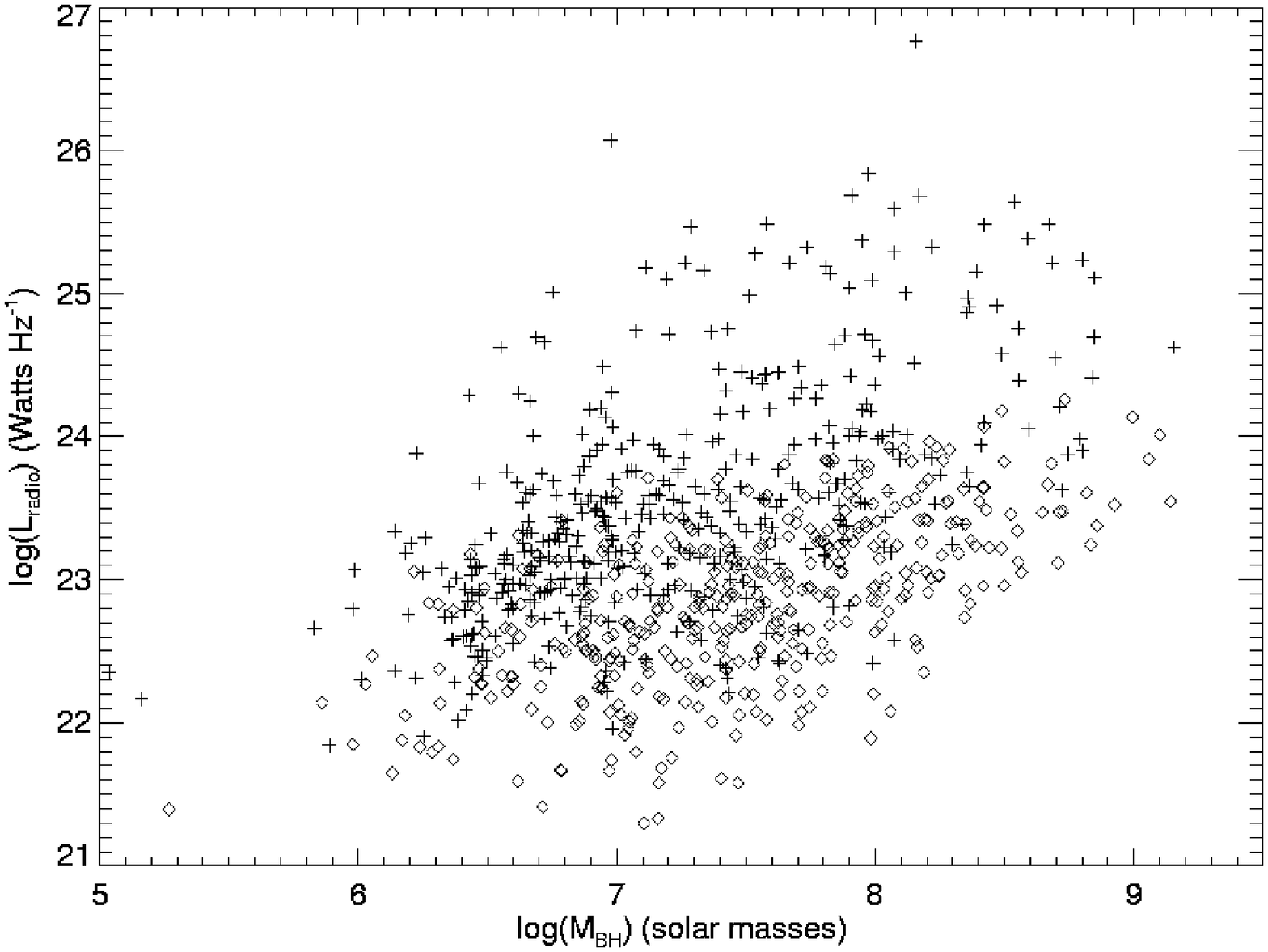}
\caption[]{}
$L_{1.4 {\rm GHz}}$ vs.\ $M_{BH}$.  The crosses have $\mathcal{R}$ $>$ 10 and the diamonds have $\mathcal{R}$ $<$ 10.
\label{fig14}
\end{figure}

\clearpage


\begin{thebibliography}

\bibitem[Becker et al.(1995)]{Becker1995} Becker, R.~H., White, 
R.~L., \& Helfand, D.~J.\ 1995, \apj, 450, 559

\bibitem[Best et al.(2005)]{Best2005} Best, P.~N., Kauffmann, 
G., Heckman, T.~M., Brinchmann, J., Charlot, S., Ivezi{\'c}, {\v Z}., 
\& White, S.~D.~M.\ 2005, \mnras, 362, 25 

\bibitem[Cirasuolo et al.(2003)]{Cirasuolo2003} Cirasuolo, M., 
Magliocchetti, M., Celotti, A., \& Danese, L.\ 2003, \mnras, 341, 993

\bibitem[Cirasuolo et al.(2004)]{Cirasuolo2004} Cirasuolo, M., 
Celotti, A., Magliocchetti, M., 
\& Danese, L.\ 2004, in AGN Physics with the Sloan Digital Sky Survey, ASP Conf.\ Ser.\ Vol.\ 311, ed.\
G.\ T.\ Richards, P.\ B.\ Hall (San Franscisco: ASP), p.\ 343

\bibitem[Greene 
\& Ho(2004)]{Greene2004} Greene, J.~E., \& Ho, L.~C.\ 2004, \apj, 610, 722

\bibitem[Greene 
\& Ho(2005)]{Greene2005} Greene, J.~E., \& Ho, L.~C.\ 2005, \apj, 630, 122

\bibitem[Greene 
\& Ho(2007)]{Greene2007} Greene, J.~E., \& Ho, L.~C.\ 2007, \apj, 667, 131

\bibitem[Ho(2002)]{Ho2002} Ho, L.~C.\ 2002, \apj, 564, 120

\bibitem[Ivezi{\'c} et al.(2002)]{Ivezic2002} Ivezi{\'c}, {\v Z}., 
et al.\ 2002, \aj, 124, 2364

\bibitem[Ivezi{\'c} et al.(2004)]{Ivezic2004} Ivezi{\'c}, Z., et 
al.\ 2004, in AGN Physics with the Sloan Digital Sky Survey, ASP Conf.\ Ser.\ Vol. 311, ed.\ G.\ T.\ Richards, P.\ B.\ Hall (San Francisco: ASP), p.\ 347

\bibitem[Jiang et al.(2007)]{Jiang2007} Jiang, L., Fan, X., 
Ivezi{\'c}, {\v Z}., Richards, G.~T., Schneider, D.~P., Strauss, M.~A., 
\& Kelly, B.~C.\ 2007, \apj, 656, 680

\bibitem[Kaspi et al.(2000)]{Kaspi2000} Kaspi, S., Smith, P.~S., 
Netzer, H., Maoz, D., Jannuzi, B.~T., \& Giveon, U.\ 2000, \apj, 533, 631

\bibitem[Kaspi et al.(2005)]{Kaspi2005} Kaspi, S., Maoz, D., 
Netzer, H., Peterson, B.~M., Vestergaard, M., 
\& Jannuzi, B.~T.\ 2005, \apj, 629, 61

\bibitem[Kellermann et al.(1989)]{Kellermann1989} Kellermann, K.~I., 
Sramek, R., Schmidt, M., Shaffer, D.~B., \& Green, R.\ 1989, \aj, 98, 1195

\bibitem[Kellermann(1999)]{Kellermann1999} Kellermann, K.~I.\ 1999, in
Active Galactic Nuclei and Related Phenomena, IAU Symp.\ 194, ed.\ Y.\ Terzian,
E.\ Khachikian, D.\ Weedman (San Fransicso: ASP), p.\ 39

\bibitem[Laor(2000)]{Laor2000} Laor, A.\ 2000, \apjl, 543, L111

\bibitem[Laor(2003)]{Laor2003} Laor, A.\ 2003, ArXiv 
Astrophysics e-prints, arXiv:astro-ph/0312417

\bibitem[McLure 
\& Dunlop(2004)]{McLure2004a} McLure, R.~J., \& Dunlop, J.~S.\ 2004, \mnras, 352, 1390

\bibitem[McLure 
\& Jarvis(2004)]{McLure2004b} McLure, R.~J., \& Jarvis, M.~J.\ 2004, \mnras, 353, L45

\bibitem[Nagar et 
al.(2005)]{Nagar2005} Nagar, N.~M., Falcke, H., \& Wilson, A.~S.\ 2005, \aap, 435, 521

\bibitem[Osterbrock 
\& Ferland(2006)]{Osterbrock2006} Osterbrock, D.~E., \& Ferland, G.~J.\ 2006, Astrophysics of gaseous nebulae and active galactic nuclei (Sausalito: University Science Books)

\bibitem[Peterson(1997)]{Peterson1997book} Peterson, B.~M.\ 1997, An 
introduction to active galactic nuclei (Cambridge: Cambridge University Press) 

\bibitem[Peterson et al.(2004)]{Peterson2004} Peterson, B.~M., et 
al.\ 2004, \apj, 613, 682

\bibitem[Schneider et al.(2007)]{Schneider2007} Schneider, D.~P., et 
al.\ 2007, \aj, 134, 102

\bibitem[Sikora et al.(2007)]{Sikora2007} Sikora, M., Stawarz, 
{\L}., \& Lasota, J.-P.\ 2007, \apj, 658, 815 

\bibitem[Spergel et al.(2003)]{Spergel2003} Spergel, D.~N., et al.\ 
2003, \apjs, 148, 175

\bibitem[Urry 
\& Padovani(1995)]{Urry1995} Urry, C.~M., \& Padovani, P.\ 1995, \pasp, 107, 803

\bibitem[Vestergaard 
\& Peterson(2006)]{Vestergaard2006} Vestergaard, M., \& Peterson, B.~M.\ 2006, \apj, 641, 689

\bibitem[White et al.(2000)]{White2000} White, R.~L., et al.\ 
2000, \apjs, 126, 133

\bibitem[White et al.(2007)]{White2007} White, R.~L., Helfand, 
D.~J., Becker, R.~H., Glikman, E., \& de Vries, W.\ 2007, \apj, 654, 99 

\bibitem[Woo 
\& Urry(2002)]{Woo2002} Woo, J.-H., \& Urry, C.~M.\ 2002, \apjl, 581, L5

\end{thebibliography}
\end{document}